\documentclass[useAMS,usenatbib]{mn2e}
\usepackage{graphicx}

\title[] {Extremely efficient Zevatron in rotating AGN magnetospheres.}

\author[Z. Osmanov, S. Mahajan, G. Machabeli and N. Chkheidze]{Z. Osmanov$^{1}$\thanks{E-mail:
z.osmanov@astro-ge.org (ZO)} S. Mahajan$^{2}$ G. Machabeli$^{3}$ and
N. Chkheidze$^{3}$
\\ $^{1}$School of Physics, Free University of Tbilisi, 0183, Tbilisi,
Georgia\\
$^{2}$Institute for Fusion Studies, The University of Texas at
Austin, Austin, TX 78712, USA\\
$^{3}$Centre for Theoretical Astrophysics, ITP, Ilia State
University, 0162 Tbilisi, Georgia}

\begin{document}

\pagerange{\pageref{firstpage}--\pageref{lastpage}} \pubyear{2011}

\maketitle

\label{firstpage}

\begin{abstract}
A novel model of particle acceleration in the magnetospheres of
rotating active galactic nuclei (AGN) is constructed.The particle
energies  may be boosted up to $10^{21}$eV in a two step mechanism:
In the first stage, the Langmuir waves are centrifugally excited and
amplified by means of a parametric process that efficiently pumps
rotational energy to excite electrostatic fields. In the second
stage, the electrostatic energy is transferred to particle kinetic
energy via Landau damping made possible by rapid  "Langmuir
collapse".  The time-scale for parametric pumping of Langmuir waves
turns out to be small compared to the kinematic time-scale,
indicating high efficiency of the first process.  The second process
of "Langmuir collapse" - the creation of caverns or low density
regions - also happens rapidly for the characteristic parameters of
the  AGN magnetosphere. The Langmuir collapse creates appropriate
conditions for transferring electric energy to boost up already high
particle energies to much higher values. It is further shown that
various energy loss mechanism are relatively weak, and do not impose
any significant constraints on maximum achievable energies.

\end{abstract}


\section{Introduction}

New observations confirming a strong correlation of  ultra-high
energy cosmic ray protons with the active galactic nuclei (AGN)
\citep{corel1} has stimulated the search for a possible mechanism
that could accelerate protons up to tens of EeV and higher energies.

In this paper we will explore energy transfer via the relativistic
centrifugal force in the rotating magnetospheres of compact objects
as a mechanism for boosting proton energies to such enormous levels.
In general, this is not a new idea, \cite{bogoval1} and
\cite{bogoval2} have considered the role of magnetocentrifugal
effects in astrophysical outflows and nonaxisymmetric
magnetospheres. Based on a rather different approach the mechanism
of centrifugal acceleration was applied to explain the observed high
and very high energy emission pattern \citep{osm7,newast} in some
astrophysical settings.

Relativistic centrifugal force turns out  to be a crucial physical
element governing particle acceleration, particularly, in
astrophysical objects endowed with rotating magnetospheres, where
magnetic energy density is larger than that of  the plasma. The
plasma particles are, thus, forced to follow the field lines, which
in turn are corotating. Consequently, the particles, sliding along
these magnetic field lines, experience centrifugal force and gain
energy.

The time-varying  centrifugal force can also parametrically excite
electrostatic modes providing an additional channel for transferring
energy from the rotator to the electromagnetic waves and , then,
possibly to the particles; this channel may be particularly relevant
to the enormously energetic AGN magnetosphere \citep{incr,incr3}.

The energy pumping mechanism  has already been applied to both
pulsars \citep{incr} and the AGN \citep{incr3}. In the AGN
magnetosphere, the centrifugally excited electrostatic waves are
found to be unstable; the growth rates, exceeding the corresponding
rate of the accretion disc evolution, indicate the possibility of an
efficient instability \citep{incr3}.

Till now, our most extensive exploration of the parametrically
driven Langmuir waves and their role in accelerating particles has
been in the context of the Crab nebula \citep{incr,screp}. We showed
that the process of generation of the centrifugally driven Langmuir
waves is rather efficient, and for the Crab pulsar magnetosphere,
energy gain of electrostatic modes might be in a reasonable
coincidence with its observed slow down rate. These waves, sustained
by the main electron-positron (magnetospheric) plasma, were shown to
Landau damp effectively on the primary electron beam ( much less
density than the main plasma), imparting large per particle energy
to the already fast particles. We have argued that the Crab pulsar,
with its extremely high rate of rotation, may guarantee acceleration
of electrons to hundreds of TeV and even higher.

In this paper, we go back to reexamine the problem begun in
\citep{incr3}; we will investigate  the efficiency of centrifugally
driven electrostatic waves in the AGN magnetosphere, and determine
if these waves can be a conduit for transferring  energy towards
particle acceleration. In contrast to the pulsars, the AGN
atmosphere contains a quasi-neutral plasmas composed of protons and
electrons, Consequently the mass dependent centrifugal force acting
will produce an initial charge separation on electrons and protons,
leading, perhaps, by means of the parametric instability, to
extremely efficient energy pumping from the rotating magnetosphere.
In addition to demonstrating the existence of unstable Langmuir
waves , we also explore a mechanism that might be responsible for
particle acceleration.

It will be helpful to remember that  the Langmuir turbulence
inevitably leads to (nonlinear) instability, creating caverns-low
density regions \citep{zakharov} . According to the investigation,
the high frequency contribution of pressure inside the cavern pulls
the particles from this area, provoking an explosive collapse of the
cavern, efficiently transferring energy from the electrostatic waves
to the particles pushed from the collapsing regions. Dynamics of
collapse was studied in detail in \citep{galeev} and numerically
investigated by \cite{degtiarev}. \cite{galeev} considered a theory
of three dimensional problem and investigated the corresponding
spectra of Langmuir turbulence, studying the role of absorption
mechanisms for short wavelength plasmons. \cite{degtiarev} analysed
the dynamics of modulational instability for long-wavelength
electrostatic oscillations and numerically studied the role of
Langmuir damping in the termination of collapse.

After spelling out our theoretical model in Sec.2, we work out  in
Sec.3 the details of the particle acceleration mechanism for typical
parameters of AGN magnetospheres. In Sec. 4, we discuss and
summarize our results.

\section[]{Theoretical model}

Transferring  the approach of \cite{screp} to the AGN atmosphere, we
divide the problem into two subtasks: (a) generation of
electrostatic waves, and (b) Particle acceleration which, in this
case, happens through the Langmuir collapse.

\subsection[]{Centrifugal excitation of Langmuir waves}

For the mode calculations of this paper, the magnetic field lines
are almost straight and co-rotate with the supermassive black hole.
For a typical AGN with mass  $M\sim 10^8\times M_{\odot}$, the
angular velocity of rotation is given by
\begin{equation}
\label{rotat} \Omega\approx\frac{a c^3}{GM}\approx
10^{-3}\frac{a}{M_8}rad/s^2,
\end{equation}
where $M_8\equiv M/(10^8M_{\odot})$ and $0<a\leq 1$ is a
dimensionless parameter characterizing the rate of rotation of the
black hole. The magnetic field  may be estimated by assuming that
the magnetic  energy density is of the order of radiation energy
density of the AGN (equipartition) \citep{osm7},
\begin{equation}
\label{mag} B\approx\sqrt{\frac{2L}{r^2c}}\approx
27.5\times\left(\frac{L}{10^{42}erg/s}\right)^{1/2}\times\frac{R_{lc}}{r}G,
\end{equation}
where $L$ is the bolometric luminosity of AGN, $R_{lc}=c/\Omega$ is
the light cylinder radius and $r$ is the distance from the black
hole. With these fields, the protons with high Lorentz factors,
$\gamma\sim 10^{4-5}$, will have gyroradii much smaller than the
kinematic length-scale ($R_{lc}$) of particles in the magnetosphere.
Such particles will, clearly, follow and co-rotate with the magnetic
field lines. The protons will, therefore, inevitably accelerate
centrifugally, but as it is explained in detail by \cite{osm7}, this
process is strongly limited by two major factors.

In due course of motion protons gain energy, but this process lasts
until the energy gain is balanced by energy losses due to the
inverse Compton scattering.The maximum attainable proton Lorentz
factor is, then,  estimated as \citep{osm7}
\begin{equation}
\label{gic} \gamma_{_{IC}}\approx\left(\frac{6\pi m_pc^4}{\sigma_T
L\Omega}\right)^2\approx 1.5\times
10^{11}\left(\frac{10^{42}erg/s}{L}\right)^2,
\end{equation}
where $m_p\approx 1.67\times 10^{-24}$g is proton's mass and
$\sigma_T\approx 6.65\times 10^{-25}$cm$^{-2}$ is the Thomson
cross-section. We have taken into account that the Black hole
rotates with $10\%$ of its maximum rate ($a = 0.1$).

This estimate, however, turns out to be  a bit too optimistic. A
more restrictive limit is set by the proton dynamics as it moves
under the combined influence of the Coriolis and Lorentz forces.
This limiting mechanism, called, for want of a better name,
"breakdown of the bead on the wire (BBW) approximation
\citep{rm00}", yields a more moderate gamma (see \citep{osm7} for
details)
\begin{equation}
\label{gfb}
\gamma_{_{BBW}}\approx\frac{1}{c}\left(\frac{e^2L}{2m_p}\right)^{1/3}\approx
1.4\times 10^{5}\left(\frac{L}{10^{42}erg/s}\right)^{1/3},
\end{equation}

%

From Eqs. (\ref{gic},\ref{gfb}), it is clear that
$\gamma_{_{BBW}}\ll\gamma_{_{IC}}$. BWW, thus, sets the theoretical
upper limit on the Lorentz factor $\sim 1.4\times 10^5$. Via
centrifugal acceleration, the energy is pumped from the rotating
magnetosphere. Although the  centrifugally accelerated protons are
strongly limited in energy, the centrifugal mechanism of energy
pumping is  very much in action.

To study the generation of extremely unstable Langmuir waves, we
follow the method developed in \citep{incr,incr3}. Throughout  the
paper, we assume that the magnetic field lines are almost straight
and lie in the equatorial plane. Then, in the $1+1$ formalism
\citep{membran}, the Euler equation in the corotating frame can be
written as
\begin{equation}
\label{eul1} \frac{d{\bf p}_{\beta}}{d\tau}= \gamma_{\beta}{\bf
g}+\frac{e_{\beta}}{m_\beta}\left({\bf E}+ \frac{1}{c}\bf
v_{\beta}\times\bf B\right),
\end{equation}
${\beta}$ denotes the species index ( electrons and protons), ${\bf
p}_{\beta}$ and ${\bf V}_{\beta}$ are, respectively,  the
dimensionless momentum (${\bf p}_{\beta}\rightarrow {\bf
p}_{\beta}/m_\beta$) and velocity, ${\bf g}\equiv
-{\bf\nabla}\xi/\xi$ is the gravitational potential (a crucial
component of the model), $e_{\beta}(m_{\beta})$ is the charge(mass)
of the corresponding particle, $d\tau\equiv\xi dt$,
$\xi\equiv\left(1-\Omega^2r^2/c^2\right)^{1/2}$ and
$\gamma_{\beta}\equiv\left(1-{\bf V}_{\beta}^2/c^2\right)^{-1/2}$ is
the Lorentz factor.

Transition to the laboratory frame (LF) is accomplished through the
identity $d/d\tau\equiv\partial/(\xi\partial t)+\left({\bf
v_{\beta}\nabla}\right)$, and the relation  $\gamma=\xi\gamma'$
connecting the corotating and LFs of reference ( the prime denotes
the quantity in the LF). The LF equation of motion
\begin{eqnarray}
\label{eul2} \frac{\partial{\bf p}_{\beta}}{\partial t}+({\bf
v_{\beta}\cdot\nabla)p}_{\beta}= \nonumber \\
=-c^2\gamma_{\beta}\xi{\bf\nabla}\xi+\frac{e_{\beta}}{m_\beta}\left({\bf
E}+ \frac{1}{c}\bf v_{\beta}\times\bf B\right),
\end{eqnarray}
coupled with the continuity
\begin{equation}
\label{cont} \frac{\partial n_{\beta}}{\partial t}+{\bf
\nabla}\cdot\left(n_{\beta}{\bf v_{\beta}}\right)=0,
\end{equation}
and the Poisson equation,
\begin{equation}
\label{pois} {\bf \nabla\cdot E}=4\pi\sum_{\beta}e_{\beta}n_{\beta}.
\end{equation}
completes the system.

In the equilibrium, the plasma is assumed to obey the frozen-in
condition ${\bf E_0 + \frac{1}{c}v_{0\beta}\times B_0 = 0}$. The
corresponding trajectory of the centrifugally accelerated particles,
calculated by the standard single particle approach, turns out to
be: $r_\beta(t) \approx \frac{V_{0\beta}}{\Omega}\sin\left(\Omega t
+ \phi_{\beta}\right)$ (see Appendix, Eq. (\ref{rtrel})) and
$\upsilon_{0\beta}(t) \approx V_{0\beta}\cos\left(\Omega t +
\phi_{\beta}\right)$, where we have taken into account the specific
initial phases of particles, $\phi_{\beta}$. It is worth noting that
in spite of the aforementioned behaviour of the radial coordinate,
it does not mean that the particle will oscillate. As we will see
later, the instability becomes so efficient that the corresponding
time-scale of energy conversion will be less than the rotation
period.

Since different species experience different forces, the Euler
equation will lead to spacial separation of charges, which in turn,
generates a electrostatic field via the Poisson equation;  under
certain conditions, these fields  may grow in time. For studying the
development of Langmuir waves, we expand all physical quantities
around this equilibrium state (keeping up to linear terms in
perturbations):
\begin{equation}
\label{exp} \Psi\approx\Psi^0+\Psi^1,
\end{equation}
where $\Psi = \{n,{\bf v},{\bf p},{\bf E},{\bf B}\}$. Fourier
decomposing the perturbations,
\begin{equation}
\label{pert} \Psi^1(t,{\bf r})\propto\Psi^1(t)exp[i{\bf kr}],
\end{equation}
 one obtains the linearized system of equations (\ref{eul2}-\ref{cont})
 \begin{equation}
\label{eul3} \frac{\partial p_{\beta}}{\partial
t}+ik\upsilon_{\beta0}p_{\beta}=
\upsilon_{\beta0}\Omega^2r_{\beta}p_{\beta}+\frac{e_{\beta}}{m_{\beta}}E,
\end{equation}
\begin{equation}
\label{cont1} \frac{\partial n_{\beta}}{\partial
t}+ik\upsilon_{{\beta}0}n_{\beta}, +
ikn_{{\beta}0}\upsilon_{\beta}=0
\end{equation}
\begin{equation}
\label{pois1} ikE=4\pi\sum_{\beta}n_{\beta0}e_{\beta},
\end{equation}
describing the evolution of the electrostatic field.

In terms of an effective density, defined by
\begin{equation}
\label{ansatz}
n_{\beta}=N_{\beta}e^{-\frac{iV_{\beta}k}{\Omega}\sin\left(\Omega t
+ \phi_{\beta}\right)},
\end{equation}
one can derive, after straightforward algebra, the following set of
non-autonomous "Mode" equations \citep{screp}
\begin{equation}
\label{ME1} \frac{d^2N_p}{dt^2}+{\omega_p}^2 N_p= -{\omega_p}^2 N_e
e^{i \chi},
\end{equation}
\begin{equation}
\label{ME2} \frac{d^2N_e}{dt^2}+{\omega_e}^2 N_e= -{\omega_e}^2 N_p
e^{-i \chi},
\end{equation}
where $\chi = b\cos\left(\Omega t+\phi_{+}\right)$, $b =
\frac{2ck}{\Omega}\sin\phi_{-}$, $2\phi_{\pm} = \phi_p\pm\phi_e$ and
$\omega_{e,p}\equiv\sqrt{4\pi e^2n_{e,p}/m\gamma_{e,p}^3}$ and
$\gamma_{e,p}$ are the relativistic plasma frequencies and the
Lorentz factors for the stream components. Note that for simplicity,
we have assumed the particle population to consist of only two
streams with initial phases $\phi_{\pm}$.

After Fourier transforming Eqs. (\ref{ME1},\ref{ME2}) in time,  we
derive the "dispersion relation" of the corresponding modes
\begin{equation}
\label{disp} \omega^2 -\omega_e^2 - \omega_p^2  J_0^2(b)= \omega_p^2
\sum_{\mu} J_{\mu}^{2}(b) \frac{\omega^2}{(\omega-\mu\Omega)^2},
\end{equation}
where $J_{\mu}(x)$ is the Bessel function. Referring the reader to
\citep{screp} for details, we just state here that this system may
undergo an instability  when the real part of the frequency
satisfies the resonance condition, $\omega_r = \mu_{res}\Omega$.
After expressing the frequency, $\omega = \omega_r+\Delta$ one can
straightforwardly reduce the above equation
 \begin{equation}
 \label{disp1}
 \Delta^3=\frac{\omega_r {\omega_p}^2 {J_{\mu_{res}}(b)}^2}{2},
 \end{equation}
that has a pair of complex solutions implying a growth rate (
imaginary part of $\Delta$)
\begin{equation}
 \label{grow}
 \Gamma= \frac{\sqrt3}{2}\left (\frac{\omega_e {\omega_p}^2}{2}\right)^{\frac{1}{3}}
 {J_{\mu_{res}}(b)}^{\frac{2}{3}},
\end{equation}
where $\omega_r = \omega_e$. The time dependent centrifugal force
that parametrically drives the electrostatic waves is different for
the two species- so are their Lorentz factors.

Due to the presence of the Bessel function, which for large values
of  the index( $\mu_{res}=\omega_r/\Omega\gg1$ tends to be nonzero
only when the argument $b$ is of the order of the index, the growth
rate will peak when $b= {2ck}\sin\phi_{-}/
{\Omega}=\mu_{res}=\omega_r/\Omega$. For these most unstable modes,
the phase velocity $\upsilon_{ph}
\equiv\omega/k=2c\sin(\varphi_{-})$ will exceed the speed of light
for certain values of $\varphi_{-}$. Since there  no particles with
such velocities, such waves will not Landau damp on the particles.
However,  if there is a process  that can enhance the wave vector,
Landau damping might be restored. it is in this context that we
discuss, in the next  subsection, the possibility of Langmuir
collapse. The length scale of a "cavern" might significantly
decrease, leading to a decrease of phase velocity, so that there are
enough resonant particles for drawing energy from the electrostatic
field.

\subsection[]{Acceleration mechanism}

It is worth noting that due to a small initial amplitude of the
Langmuir wave, the high frequency pressure increases, pushing out
the particles from the perturbed zone. The resulting  polarization
creates an additional electrostatic field which causes a further
decrease density in this region, creating what are termed as
"caverns". Simultaneously, the plasmons (quasiparticles) accelerate
towards these cavities, enter them and enlarge their depth even
more. The  boosted high-frequency pressure induces an auto
modulation instability of the spatial distribution of plasmons.

For the purpose of this paper, we will consider a kinematically
relativistic fluid with  non-relativistic temperatures. In the rest
frame, it has been shown that
 the fluid obeys the nonlinear set of hydrodynamic equations \citep{zakharov},
 \begin{equation}
 \label{z1}
 \frac{\partial}{\partial t}\delta n+n_0 \nabla{\bf v_e} = 0,
 \end{equation}
 \begin{equation}
 \label{z2}
 \frac{\partial}{\partial t}{\bf v_e}+\frac{e}{m}\nabla{\varphi_e}+\frac{3}{2}\frac{T_e}{2mn_0}\nabla{\delta n} = 0,
 \end{equation}
reduces to
\begin{equation}
\label{ez1} \left[\frac{\partial^2}{\partial
t^2}-3\lambda_D^2\omega_p^2\frac{\partial^2}{\partial
x^2}-\omega_p^2\right]E = \frac{\delta n}{n_0}\omega_p^2E,
\end{equation}
\begin{equation}
\label{ez2} \left[\frac{\partial^2}{\partial
t^2}-\lambda_D^2\omega_p^2\frac{\partial^2}{\partial
x^2}\right]\delta n=\frac{1}{16\pi m_p}\frac{\partial^2E^2}{\partial
x^2},
\end{equation}
where $\delta n$ is the electron density perturbation, ${\bf v_e}$
is the corresponding velocity perturbation, $n_0$ is the unperturbed
ion number density, $T_e$ is the electron temperature, $\varphi_e$
is the high frequency part of the electrostatic potential and
$\lambda_D\equiv \sqrt{T_e/(4\pi n_0e^2)}$ is the Debye length
scale. We assume that the plasma is quasi neutral.

It has been shown that this system, generalized for higher
dimensions, describing the Langmuir collapse, exhibits explosive
behaviour, i.e,   the initial growth rate of the instability
\citep{arcimovich}
\begin{equation}
\label{incrcol}
\Gamma_{LC}=Im(\omega)\approx\frac{1}{\gamma_p}\left[\frac{\langle
E^2\rangle e^2m_p}{4k_BT}\right]^{1/2},
\end{equation}
since it scales with the field intensity, will naturally increase
with time. In deriving (\ref{incrcol}), we have taken into account
the Lorentz boosting. According to the standard approach developed
by \cite{zakharov}, kinetic and potential energies of the plasmons
caught by the caverns are of the same order of magnitude
\begin{equation}
\label{k} k^2\lambda_D^2\sim\frac{\mid\delta n\mid}{n_0}.
\end{equation}
%


Thus the characteristic length-scale of the cavern $L_c$, defined by
the  wavelength of the plasmon($1/k$), scales inversely with density
perturbation, i.e, $L_c~ 1/\sqrt{\delta n}$. Since the high
frequency pressure, $P_{hf}=-|E|^2\delta n/(24k^2\lambda_D^2n_0)$
\citep{arcimovich} with $|\delta n|$, the greater the pressure the
lesser the cavern width. This is precisely the recipe for an
explosive instability leading to a smaller and smaller $L_c$. During
this process, energy density of oscillations drastically increases
accelerating the collapse.  What happens later depends strongly on
dimensionality of the process.
 If the collapse continues so that the $L_c=1/k$
when the wavelength of plasmons is of the order of the Debye scale,
the resonance Landau damping becomes important, resulting in
particle acceleration. More  discussion follows in the next section.

\section{Discussion}
In this section we consider the implications of the theoretical
framework developed in the last section (instability of
centrifugally induced electrostatic waves, and the expected Langmuir
collapse) for particle acceleration in a typical AGN setting.

It has been assumed that the magnetic field are so strong that
inside the light cylinder zone the AGN magnetospheric plasma
co-rotates rigidly. Therefore, in this area the plasma number
density might be well approximated by the Goldreich-Julian density
\begin{equation}
\label{gj} n_0 = \frac{\Omega B}{2\pi ec}.
\end{equation}
If we assume that the maximum electron Lorentz factor is  controlled
by the same mechanism as protons ($\gamma_p\sim 1.4\times 10^5$),
then their lighter mass will allow
$\gamma_e\approx\gamma_p(m_p/m_e)^{1/3}\sim 1.6\times 10^6$ (see
equation (\ref{gfb})) to be an order of magnitude larger.
Considering two representative streams with  Lorentz factors
$\gamma_1\sim 1.6\times 10^6$ and $\gamma_2\sim 10^3$ and tempera
ture $T\sim 10^4$K, one estimates that the instability time-scale,
$1/\Gamma$, is less than the kinematic time-scale, $\sim
2\pi/\Omega$, implying  a rather efficient process  of pumping
rotation energy into Langmuir waves. These waves will, then,
accelerate particles by Landau damping aided by a possible Langmuir
collapse.

Since the perturbation of density modulation usually is small,
$\delta n\ll n_0$, the change of frequency of plasmons is
negligible, $\delta\omega\ll\omega$. One may assume, then, that the
corresponding energy is constant \citep{arcimovich}
\begin{equation}
\label{E2a} \int d^q{\bf r}\mid E\mid^2 = const.
\end{equation}
implying the scaling
\begin{equation}
\label{E2} \mid E\mid^2\propto\frac{1}{l^q},
\end{equation}
where $q$ equals $\{1,2,3\}$ depending on the dimension of the
space. Inside the magnetosphere the magnetic field is strong enough
to force particles to move along the field lines. The relevant
geometry, thus, is 1 D and the high frequency pressure,
$P\propto\mid E\mid^2$ scales as $1/l$  \citep{arcimovich}. On the
other hand, the thermal pressure, $P_{th}=k_BT\delta n$, behaving as
$1/l^2$ (see equation (\ref{k})) increases faster than the high
frequency pressure. This means that inside the magnetosphere the
centrifugally induced Langmuir waves do not collapse and only
propagate towards the outer region of the magnetosphere.

Outside the magnetosphere, the plasma kinematics are no longer
governed by rotation. in this region, the density is controlled by
the accretion processes. To estimate the density let us assume a
spherically symmetric accretion. The expected  accretion rate of
particles per unit time and unit  area is of the order of
$n\upsilon$, where $n$ is the accretion particle number density and
$\upsilon = \sqrt{GM_{_{BH}}/R_{lc}}$. The estimated density, then,
is
\begin{equation}
\label{n} n=\frac{L}{4\eta\pi m_pc^2\upsilon R_{lc}^2}\approx
6.3\times 10^5\times \left(\frac{L}{10^{42}erg/s}\right) cm^{-3}.
\end{equation}
We have assumed that only $10\%$ of the rest energy of accretion
matter ($\eta=0.1$) transforms to emission. For such dense matter
the corresponding plasma frequency, $\omega_p=\sqrt{4\pi e^2n/m_p}$
exceeds the cyclotron frequency for protons
$\omega_B=eB_{lc}/(m_pc)$, which means that the particles in the
region outside the magnetosphere are no longer bound by the magnetic
field and consequently $d = 3$. This in turn means that the high
frequency pressure behaving as $1/l^3$, increases much faster than
the thermal pressure, behaving as $1/l^2$ and correspondingly the
collapse of the Langmuir waves becomes inevitable.

During the collapse the density perturbation satisfies the
approximate equation (Eqs. (\ref{ez1},\ref{k}))
\begin{equation}
\label{dnt} \frac{\partial^2\delta n}{\partial
t^2}\approx\frac{\delta n\mid E\mid^2}{16\pi nm_p\lambda_D^2},
\end{equation}
where the thermal pressure has been neglected.

After complementing this equation with already discussed relations,
$\mid E\mid^2\sim 1/l^3$ and $\delta n\sim 1/l^2$, one obtains
 \citep{zakharov}
\begin{equation}
\label{E2} \mid E\mid^2\approx \mid
E_0\mid^2\left(\frac{t_0}{t_0-t}\right)^2
\end{equation}
\begin{equation}
\label{l} l\approx l_0\left(\frac{t_0}{t_0-t}\right)^{-2/3},
\end{equation}
where $t_0$ is the  time when the cavern collapses completely.

It is worth noting that the initial Langmuir waves have been
efficiently amplified in the very vicinity of the light cylinder
surface. The corresponding lengthscale can be obtained by a simple
approximate expression $\Delta r\approx\gamma/(d\gamma/dr)$, leading
to
\begin{equation}
\label{dr} \Delta r\approx\frac{\gamma_0}{2\gamma}R_{lc},
\end{equation}
where we have taken into account the radial behaviour of Lorentz
factors of centrifugally accelerated particles \citep{rm00}
\begin{equation}
\label{gr} \gamma(r)=\frac{\gamma_0}{1-\frac{r^2}{R_{lc}^2}}.
\end{equation}
As we have already discussed the electrostatic field appears and
amplifies by means of the separation of charges in the mentioned
zone. Therefore, the electrostatic field is approximated by the
Poisson equation
\begin{equation}
\label{E0} E_0\approx 4\pi ne\Delta r,
\end{equation}
which due to the Langmuir collapse will be boosted by the factor of
$\left(\Delta r/l\right)^{3/2}$ (see Eqs.(\ref{E2},\ref{l})) where
$l\approx 2\pi\lambda_D$ is the dissipation lengthscale
\citep{arcimovich}. It is clear that at the final stage, energy of
the amplified electrostatic field will transfer to the particles
inside the cavern resulting in protons with extremely high energies:
\begin{equation}
\label{energy} \epsilon_p\approx\frac{E^2}{8\pi
n}=\frac{ne^2}{4\pi^2\lambda_D^3}\Delta r^5.
\end{equation}
For a proton beam with $\gamma_p\sim 10^2$, one can easily calculate
that even the initial instability time-scale measured in the lab
frame $\sim 1/\Gamma_{LC}\sim 0.2$s is several orders of magnitude
less than the kinematic time-scale. This difference will
significantly increase, because the collapse has an explosive
character and the electric field amplifies faster than the linear
exponential increase. The Langmuir collapse is strong enough to
guarantee efficient acceleration of particles to ultra high
energies. In particular, from equation (\ref{energy}), one obtains
\begin{equation}
\label{energy1} \epsilon_p\left(eV\right)\approx 6.4\times
10^{17}\times\left(\frac{f}{10^{-3}}\right)^3\times\left(\frac{10^2}{\gamma_2}\right)^5
\times M_8^{-5/2}\times L_{42}^{5/2},
\end{equation}
where $f = \delta n/n_0$ is the initial dimensionless density
perturbation, $a = 0.1$, $\eta = 0.1$, $L_{42}\equiv L/10^{42}$erg/s
and we have assumed $T\sim 10^4$K. As it is evident from this
expression for a convenient set of parameters one can achieve
enormous  energies of the order of $10^{21}$eV. To achieve such
energies, the required electrostatic fields must exceed the
background magnetic field by many orders of magnitude (see Eqs.
(\ref{mag},\ref{E0},\ref{energy})). This is possible because the
origin of the electrostatic field is different from that of the
background magnetic field; the magnetospheric rotation energy is
almost a limitless and continuous source, and can readily feed
electric fields of such enormous magnitude. For the parameters
corresponding to the maximum attainable energy $10^{21}$eV, the
length-scale of the cavern is of the order of $10^{12}$cm (see
equation \ref{dr}). Again such a large scale structure of the
electrostatic field can be maintained only because the available
energy budget that is transferred to the Langmuir modes, is huge. It
should also be stressed  that the equilibrium frozen in condition (a
condition relating the large scale equilibrium fields $E_0$ and
$B_0$) is not affected by the much shorter scale electric fields
associated with the Langmuir wave.

During the acceleration phase, the particles might lose energy due
to several mechanisms that, potentially, might reduce the overall
efficiency of acceleration. Via the highly efficient synchrotron
mechanism, for example, the particles may rapidly lose their
perpendicular momentum, transit to the ground Landau level and slide
along the magnetic field lines. This mechanism, thus, does not
influence particle acceleration.

The Inverse Compton scattering is also found to be little
significance to the acceleration process. For such high energies,
the relevant regime is Klein-Nishina, and
 the corresponding cooling time-scale goes as
$t_{_{IC}}=\epsilon_p/P_{KN}\propto\epsilon_p$ \citep{or09}, where
$P_{KN}$ is the power emitted per unit  time; $P_{KN}$  is not
sensitive to $\epsilon_p$ \citep{Blumenthal}. Since $t_{_{IC}}$ is a
continuously increasing function of $\epsilon_p$, the inverse
Compton process does not impose any constraints on achievable
particle energies.

Another mechanism is the curvature radiation, characterized by the
cooling time-scale $t_{cur}=\epsilon_p/P_{cur}$, where $P_{cur} =
2e^2\epsilon_p^4/(3m_p^4c^3\rho)$ is the energy loss rate and $\rho$
is the curvature radius of the trajectory of particles. Acceleration
is efficient until the acceleration time-scale, which is the
collapse time-scale, is less than the cooling time-scale. Maximum
energy is achieved when the following condition $t_{col}\approx
t_{cur}$ is satisfied. If one takes into account the gyro radius,
$R_{p}$, of relativistic protons and assumes $\rho\sim R_p$, one can
obtain $\epsilon_{max}\approx 4\times 10^{24}\xi$ (eV), where
$\xi\gg 1$ and characterizes the fact that the growth rate is much
higher than the initial increment (see equation (\ref{incrcol})).
Therefore, the curvature emission is also negligible and cannot
impose notable constraints on the maximum energies of particles.

Although we see that the present mechanism, in principle, might
create primary cosmic ray (proton) energies  in the ZeV range, the
actual energies may be limited due to the interaction of these
particles with the isotropic microwave cosmic radiation. This
interaction could significantly reduce the proton energy to the so
called GZK limit $4\times 10^{19}$eV \citep{gzk1,gzk2}. There is,
however, ample observational evidence of cosmic rays with energies
above the GZK limit \citep{overgzk}. This can happen, for instance,
if there is not enough time for the background radiation to slow
down the more energetic particles - that is- the source of highly
energetic particles lies within what may be called the  the GZK
radius, $\sim 100$Mpc \citep{corel1}- the typical distance needed
for significant energy loss on the cosmic microwave photons.

In the context of this paper it is worth noting that, the strong
correlation of AGN with cosmic rays is actively discussed by
\cite{corel1}, who consider the possibility that a subclass of AGN,
might be responsible for ultra-high energy cosmic rays.

\section{Summary}
\begin{enumerate}

      \item We have developed a new mechanism of particle
      acceleration operating in AGN magnetospheres.
      The mechanism, capable of creating extremely high energy cosmic rays,
      consists of two major stages: (I) centrifugal
      excitation of Langmuir waves and (II) the collapse of these
      waves by means of the modulational instability leading to
      particle acceleration.

     \item For studying the excitation of electrostatic
     waves parametrically induced by relativistic centrifugal force
     we considered the linearized system of equations governing this process.
     It was found that the growth time of the instability is
     small compared to the kinematic time-scale, indicating the high
     efficiency of energy pumping from rotation to Langmuir waves in
     the light cylinder zone, where relativistic centrifugal effects
     become important.

     \item As a next step we examined the Langmuir collapse of
     caverns (low density regions) by means of the high frequency
     pressure, exceeding the thermal pressure. This in turn, results in
     acceleration of protons up to ZeV energies. It has also been
     found that during acceleration major mechanisms governing energy losses
     do not impose any significant constraints on the maximum
     attainable proton  energies.

      \end{enumerate}

\section*{Acknowledgments}

The research of ZO, GM and NC was partially supported by the Shota
Rustaveli National Science Foundation grant (N31/49). The work of SM
was, in part, supported by USDOE Contract No.DE-- FG 03-96ER-54366.

\appendix

\section{}

In this section we consider a single particle, sliding along a
corotating straight wire. It is clear that in the LF of reference
the particle experiences a reaction force from the wire. Determining
the relativistic momentum of particles
\begin{equation}
\label{pr} P_r = \gamma m\frac{dr}{dt},
\end{equation}
\begin{equation}
\label{pf} P_{\varphi} = \gamma mr\Omega,
\end{equation}
where $\gamma =
\left(1-\Omega^2r^2/c^2-\upsilon^2/c^2\right)^{-1/2}$ is the Lorentz
factor and $\upsilon\equiv dr/dt$, the equations of motion in the LF
of reference are given by
\begin{equation}
\label{fr} \frac{dP_r}{dt}-\Omega P_{\varphi} = 0,
\end{equation}
\begin{equation}
\label{pf} \frac{dP_{\varphi}}{dt}+\Omega P_{r} = F,
\end{equation}
where the first equation describes dynamics of motion and the second
equation defines a value of the reaction force. We have taken into
account the relations for unit vectors in polar coordinates:
$\frac{d{\bf e_r}}{dt} = \Omega{\bf e_{\varphi}}$, $\frac{d{\bf
e_{\varphi}}}{dt} = -\Omega{\bf e_r}$ and applied the fact that the
reaction force is perpendicular to the wire: ${\bf F} = {\bf
e_{\varphi}}F$.

After quite straightforward calculations, one can show that equation
(\ref{fr}) reduces to \citep{mr94}
\begin{equation}
\label{ar} \frac{d^2r}{dt^2} =
\frac{\Omega^2r}{1-\frac{\Omega^2r^2}{c^2}}\left[1-\frac{\Omega^2r^2}{c^2}-2
\left(\frac{dr}{dt}\right)^2\right],
\end{equation}
having the following solution
\begin{equation}
\label{rt} r(t) = \frac{\upsilon_0}{\Omega}\frac{sn\left(\Omega
t\right)}{dn\left(\Omega t\right)}.
\end{equation}
Here, $\upsilon_0$ is the initial radial velocity of the particle
and $sn$ and $dn$ are a Jacobian elliptical sine and a modulus,
respectively \citep{abrsteg}. It is worth noting that for initially
relativistic particles ($\upsilon_0\approx c$) the aforementioned
expression, approximately reduces to
\begin{equation}
\label{rtrel} r(t) \approx \frac{\upsilon_0}{\Omega}\sin\left(\Omega
t\right).
\end{equation}

\bsp

\label{lastpage}


\begin{thebibliography}{99}
\bibitem[\protect\citeauthoryear{Abraham et al.}{2008}]{overgzk} Abraham, J. et al., 2008,
Phys. Rev. Lett., 101, 061101
\bibitem[\protect\citeauthoryear{Abramowitz \& Stegun}{1965}]{abrsteg} Abramowitz, M. \&
Stegun, I. A., {\it Handbook of Mathematical Functions}, edited by
Abramowitz, M. \& Stegun, I. A., Natl. Bur. Stand. Appl. Math. Ser.
No. 55 (U.S. GPO, Washington, D.C., 1965)

\bibitem[\protect\citeauthoryear{Artsimovich \& Sagdeev}{1979}]{arcimovich} Artsimovich, L.A. \&
Sagdeev, R.Z. {\it Plasma Physics for Physicists}, Atomizdat, Moscow
\bibitem[\protect\citeauthoryear{Blumenthal \&
Gould}{1970}]{Blumenthal} Blumenthal, G.R. \& Gould, R.J., 1970,
Rev. Mod. Phys., 42, 237
\bibitem[\protect\citeauthoryear{Bogovalov}{2001}]{bogoval2}
Bogovalov, S., 2001, A\&A, 367, 159
\bibitem[\protect\citeauthoryear{Bogovalov \& Tsinganos}{1999}]{bogoval1}
Bogovalov, S. \& Tsinganos, K., 1999, MNRAS, 305, 211
\bibitem[\protect\citeauthoryear{Degtiarev et al.}{1976}]{degtiarev}
Degtiarev, L.M., Zakharov, V.E. \& Rudakov, L.I., 1976, Sov. J.
Plasma Phys., 2, 240
\bibitem[\protect\citeauthoryear{Galeev et al.}{1977}]{galeev} Galeev, A.A.,
Sagdeev, R.Z., Shapiro, V.D. \& Shevchenko, V.I., 1977, Sov. J. Exp.
Theor. Phys. Lett., 46, 711


\bibitem[\protect\citeauthoryear{Greisen}{1966}]{gzk2} Greisen, K., 1966, Phys. Rev. Lett., 16, 748

\bibitem[\protect\citeauthoryear{Kim \& Kim}{2013}]{corel1} Kim, Hang Bae \& Kim, Jihyun, 1013,
Int. J. Mod. Phys. D, 22, 1350045


\bibitem[\protect\citeauthoryear{Machabeli \& Rogava}{1994}]{mr94} Machabeli, G.Z. \& Rogava, A.D.,
1994, Phys. Rev. A 50, 98
\bibitem[\protect\citeauthoryear{Mahajan et al.}{2013}]{screp}
Mahajan, S., Machabeli, G., Osmanov, Z. \& Chkheidze, N., 2013, Nat.
Sci. Rep. 3, 1262
\bibitem[\protect\citeauthoryear{Machabeli et al.}{2005}]{incr} Machabeli, G.,
Osmanov Z. \& Mahajan, S., 2005, Phys. Plasmas 12, 062901

\bibitem[\protect\citeauthoryear{Osmanov }{2010}]{newast} Osmanov,
Z., 2010, New. Astron., 15, 351
\bibitem[\protect\citeauthoryear{Osmanov }{2008}]{incr3} Osmanov,
Z., 2008, Phys. Plasmas, 15, 032901
\bibitem[\protect\citeauthoryear{Osmanov \& Rieger}{2009}]{or09}
Osmanov, Z. \& Rieger, F.M., 2009, A\&A, 502, 15
\bibitem[\protect\citeauthoryear{Osmanov et al.}{2007}]{osm7}
Osmanov, Z., Rogava, A.S. \& Bodo, G., 2007, A\&A, 470, 395
\bibitem[\protect\citeauthoryear{Rieger \& Mannheim}{2000}]{rm00} Rieger, F.M. \& Mannheim, K.,
2000, A\&A, 353, 473

\bibitem[\protect\citeauthoryear{Thorne et al.}{1986}]{membran} Thorne, K., Price, R. \& MacDonald, D.A. {\it Black Holes: The
Membrane Paradigm}, Yale University Press, New Haven, CT (1986)

\bibitem[\protect\citeauthoryear{Zakharov}{1972}]{zakharov} Zakharov, V.E., 1972,
Sov. J. Exp. Theor. Phys., 35, 908
\bibitem[\protect\citeauthoryear{Zatsepin \& Kuz'min}{1966}]{gzk1} Zatsepin, G.T. \& Kuz'min, V.A. 1966,
Sov. J. Exp. Theor. Phys., 4, 78





\end{thebibliography}
\end{document}